\documentclass{amsart}
\usepackage{amssymb,latexsym}

\newcommand{\ignore}[1]{}

\newcommand{\Ints}{\ensuremath{\mathbb{Z}}}   

\newcommand{\suchthat}{\; : \;}

\newtheorem{theorem}{Theorem}      
\newcommand{\BT}{\begin{theorem}}
\newcommand{\ET}{\end{theorem}}

\newtheorem{definition}{Definition}      %
\newcommand{\BD}{\begin{definition}}
\newcommand{\ED}{\end{definition}}
\newtheorem{corollary}[theorem]{Corollary}      %
\newcommand{\BCR}{\begin{corollary}}
\newcommand{\ECR}{\end{corollary}}
\newtheorem{example}{Example}
\newcommand{\BEX}{\begin{example}}
\newcommand{\EEX}{\end{example}}
\newtheorem{lemma}{Lemma}[theorem]  
\newcommand{\BL}{\begin{lemma}}
\newcommand{\EL}{\end{lemma}}
\newtheorem{proposition}[theorem]{Proposition}
\newcommand{\BP}{\begin{proposition}}
\newcommand{\EP}{\end{proposition}}
\newtheorem{claim}[theorem]{Claim}            %
\newcommand{\BCM}{\begin{claim}}
\newcommand{\ECM}{\end{claim}}
\newtheorem{fact}[theorem]{Fact}            %
\newcommand{\BF}{\begin{fact}}
\newcommand{\EF}{\end{fact}}

\newcommand{\BPF}{\begin{proof}}
\newcommand{\EPF}{\end{proof}}
\newcommand{\BE}{\begin{enumerate}}
\newcommand{\EE}{\end{enumerate}}
\newcommand{\BI}{\begin{itemize}}
\newcommand{\EI}{\end{itemize}}

\newcommand{\QuadSpace}{\vspace{0.25\baselineskip}}

\begin{document}

\title{An analysis of a bounded resource search puzzle}

\author{Gopal Ananthraman}
\address{Borland Software Corp.\\
         100 Enterprise Way\\
         Scotts Valley, CA 95066}
\email{gananthraman@borland.com}

\begin{abstract} 

Consider the commonly known puzzle, given $k$ glass balls, find an optimal algorithm
to determine the lowest floor of a building of $n$ floors from which a thrown glass 
ball will break. This puzzle was originally posed in its original form in \cite{focs1980}
and was later cited in the book \cite{algthc}. There are several internet sites that presents
this puzzle and its solution to the special case of $k=2$ balls. This is the first such analysis
of the puzzle in its general form. Several variations of this puzzle have been studied
with applications in Network Loading \cite{cgstctl} which analyzes a case similar to a scenario
where an adversary is changing the lowest floor with time. Although the algorithm specified in 
\cite{algthc} solves the problem, it is not an efficient algorithm. In this paper another
algorithm for the same problem is analyzed. 
It is shown that if $m$ is the minimum number of attempts required then 
for $k \geq m$ we have  $m = \log (n+1)$ and for $k < m$ we have,
$1 + \sum_{i=1}^{k}{{m-1}\choose{i}} < n \leq \sum_{i=1}^{k}{{m}\choose{i}}$ 
\end{abstract}

\maketitle  

\section{Introduction}

Consider the case $k=1$. It is trivial to note that at least $n$ sequential tries are required to determine the lowest floor,
if it exists. Note that we are considering the case where the lowest floor may or may not exist.
If we make the assumption that there is a guaranteed existence of a lowest floor, then in the 
worst case ( where $n$ itself is the lowest floor ) we need $n-1$ attempts. 
The problem becomes interesting for $k=2$. If there are $2$ balls then if a ball breaks at a floor (say , $i$)
then the other ball can be used to determine the lowest floor, which must be $\leq i$. 
This case and the more general version is analysed in the following sections.

\section{Definitions}

The following definitions are used in the rest of the document.\\
\begin{itemize}
       \item{$n$ : number of floors}
       \item{$k$ : number of glass balls available}
       \item{$m$ : Minimum number of attempts required to find $L$ using $k$ balls in the $worst case$, \\
             for a building with $n$ floors.}
 
 \item{$\{p_1, p_2, p_3,\dots,p_m\}$ : Set of floor indices where an attempt is made for $n$ floors.\\
 and finally, $p_0 = 0$ for convinience.}
\end{itemize}
 
We would like to find the sequence  $\{p_1, p_2, p_3,\dots,p_m\}$ and $m$
Given $n$ and $k$ and the assuming the existence of such a floor.

\section{Analysis for $2$ balls}
This is a more common version of the puzzle and the solution can be found from several
internet posts. We still cover this version for better insights.
For the case $k=2$ note that after the first ball breaks at whatever $p_i$ the number
of attempts for the remaining is restricted to $p_i - p_{i-1} - 1$ since there is only
one ball left and in the worst case $p_i-1$ could be $L$.
Now After an attempt at $p_i$  we have only $m - i$ attempts left. 
Note that inorder to result in $m$ number of attempts we \emph{have} to use all of $m-i$ remaining attempts.
Any less usage will imply that $m$ is not the minimum, which is a contradiction to the
initial definition of $m$.
Now, $m-i$ attempts all has to use the remaining one glass ball.So $m-i$ floors is the best 
we can do.
Now,\\
\begin{align}\label{Eq:1}
p_i - p_{i-1} - 1 = m - i
\end{align}
\begin{align}\label{Eq:2}
m + \sum_{i=1}^{m}{p_i - p_{i-1} - 1} = n  
\end{align}
The first equation is trivial. The second is obtained by simply adding the 
floors we probe and the floors between the probed floors. This should
be the total number of floors.

So from (\ref {Eq:1}) and (\ref {Eq:2}) we have,
\begin{align*}
  n &=  m + \sum_{i=1}^{m}{p_i - p_{i-1} - 1}\\
    &=  \frac{m(m+1)}2
\end{align*}
which gives, \\
\begin{align*}
  m = \frac{\sqrt {1 + 8n} -1}2
\end{align*}
Also from (\ref{Eq:1}) we can find values for $p_i$.

\section{Analysis for $k$ balls}

Note that the optimal number of attempts for $k$ balls ($m$)
is determined totally by the optimal number of attempts using $k-1$ balls.
More specifically, If a ball breaks at a probe point (Say, $p_b$) and did not
break at the last probe point ($p_g$), the remaining number of attempts must be the best possible(optimal)
for the remaining balls for sufficiently large $p_b - p_g -1$ floors.

Now,
Consider a function $\boldsymbol {P} \suchthat \Ints X \Ints \to \Ints$ s.t
\begin{align}\label{Eq:3}
 P(m-1, k) < n <= P(m, k) \\ 
 P(0, 0) = P(0, x) = P(x, 0) = 0 
\end{align}
Intuitively this means that there exists a function that determines the lowest floor in
*exactly* $m$ trials and *no less*. Here we assume that such a function exists. We validate this assumption later by explicitly
providing such a function instance.If the same function is applied for lesser than minimal(m) attempts (with same number of balls)
it will not be possible to determine the lowest floor.We are seeking a function that tight bounds $n$. Once this function
is found we can seek $m$ by fixing $n$.
By applying the optimal subsolution strategy described above (similar to dynamic programming) and from (\ref {Eq:2}) we have, \\
\begin{align}\label{Eq:4}
P(m, k) = m + \sum_{j=1}^{m-1}{P(j, k-1)}
\end{align}
Intuitively, this follows from the fact that at $p_i$ we have already expended $i$ attempts
and if a ball breaks at $p_i$, then the optimal number of attempts for the remaining floors  
must be $m-i$ with $k-1$ balls and the remaining floors ($p_i - p_{i-1} -1$) itself must be bounded by 
$P(m-i, k-1)$ from above as in (\ref {Eq:3}). Other way of looking at this is by assuming $n = P(m,k)$ case.
Now from (\ref {Eq:4}) we have 
\begin{align}\label{Eq:5}
P(m-1, k) = m-1 + \sum_{j=1}^{m-2} {P(j, k-1)}
\end{align}

Subtracting (\ref {Eq:5}) from (\ref {Eq:4}) we have,
\begin{align*}
P(m,k) = 1 + P(m-1, k-1) + P(m-1, k)
\end{align*}

Solving this recurrence yields
\begin{align}\label{Eq:6}
P(m, k)  &= \sum_{j=1}^{k}{m \choose j}
\end{align}
Note that there is no closed form for the binomial sum.

if $k = m$, then the sum becomes,\\
\begin{align*}
n  &= \sum_{j=1}^{m}{m \choose j} \\
   &= 2^{m} - 1 \\
 m &= \log(n+1)
\end{align*}

proving that familiar binary search is optimal.
For $k >m$, the terms $m \choose {m+1}$ upto $k$ becomes $0$ so $m = \log (n+1)$  still holds and 
binary search is optimal.\\
For $k < m$ and given $n$ the best we can do is to solve the polynomial in (\ref {Eq:6}) and $\lceil m \rceil$
gives the optimal attempts.\\
For calculating the $p_i$s Note that \\
\begin{align*}
p_1  &= 1 + P(m-1, k-1) \\
     &= 1 + \sum_{j=1}^{k-1}{{m-1} \choose {j}}
\end{align*}
\begin{equation}
p_2  = P(m-2, k-1) + p_1 + 1
\end{equation}
and so on.

This gives the exact values of $p_i$s. Note that the algorithm to follow to find the minimal floor 
is dependent on the number of resource available (no. of balls).

\section{conclusion}

It can be seen that the algorithm presented for the problem is the best possible, based on the
initial definitions of $m$. Several variations of this problem can be studied with cost function per probe etc. 
Combinatorial Puzzles, in general, serve as useful models for larger problems and are a good test bed
for solutions to larger problems. Applications of this puzzle to flow control is an example.

\end{document}